\documentstyle[12pt,epsfig]{article}
\begin{document}

\title{$V_{cs}$ from Pure Leptonic Decays of $D_s$ with Radiative Corrections}

\author{Guo-Li Wang$^{a,c}$, Tai-Fu Feng$^{b,c}$ and Chao-Hsi Chang$^c$  }
\maketitle
\begin{center}
$a$, Department of Physics, FuJian Normal University,
 FuZhou 350007, China\\
$b$, Department of Physics, NanKai University,
 TianJin 300070, China\\
$c$, Institute of Theoretical Physics, Academia Sinica,
 P.O.Box 2735, BeiJing 100080,China\\

 \end{center}
 \baselineskip 24pt
\small{
\begin{center}
\begin{abstract}
The radiative corrections to the pure 
leptonic decay $D_s{\longrightarrow} 
{\ell}{ {\nu}}_{\ell}$ up-to one-loop order 
is presented.  We find the virtual photon
 loop corrections to $D_s{\longrightarrow} 
{\tau}{ {\nu}}_{\tau}$ is negative and the corresponding
branching ratio is larger than $3.51\times 10^{-3}$. Considering the
possible experimental resolutions,
 our prediction of the
radiative decay $D_s{\longrightarrow} 
{\tau}{ {\nu}}_{\tau}\gamma$ is not so large as others, and
the best radiative channel to determine the $V_{cs}$ or $f_{D_s}$ is
 $D_s{\longrightarrow} 
{\mu}{{\nu}}_{\mu}{\gamma}$.

\end{abstract}
\end{center}}
\small{PACS numbers:{\bf 13.20.Fc, 12.39.Jh, 13.40.Ks}}

\indent
   
The pure-leptonic 
decay $D_s{\longrightarrow} {\ell}{{\nu}_{\ell}}$ 
can be used to determine the decay constant $f_{D_s}$
if the fundamental Cabibbo-Kobayashi-Maskawa matrix element $V_{cs}$ of 
Standard Model (SM) is known. Conversely if we know the value of decay
constant $f_{D_s}$ from other method\cite{as,cr,cw}, these process also can 
be used to extract the matrix element $V_{cs}$. But there are the well 
known effect of helicity suppression we can see it 
by factor of $m_{\ell}^{2}/m_{D_s}^{2}$:
\begin{equation}
\Gamma(D_s{\longrightarrow}{\ell}{\overline{\nu}_{\ell}})=
{\frac{G_{F}^{2}}{8\pi}}|V_{cs}|^{2}f_{D_s}^{2}m_{D_s}^{3}{\frac
{{m_{\ell}}^{2}}{{m_{D_s}}^{2}}}
\left(1-{\frac{{m_{\ell}}^{2}}{{m_{D_s}}^{2}}}\right)^{2},
\end{equation}
Of them only the process $D_s{\longrightarrow} {{\tau}}{\nu}_{\tau}$ 
is special, it does not suffer so much 
from the helicity suppression, and its branching ratio may reach
to $4.5\%$ in SM. However the 
produced $\tau$ will decay promptly and one more neutrino  
is generated in the cascade decay at least, thus it makes 
the decay channel difficult to be observed.
 For the channels $D_s{\longrightarrow} {{e}}{\nu}_{e}$ and 
 $D_s{\longrightarrow} {{\mu}}{\nu}_{\mu}$, 
 besides the small branching ratios, there are only
one detected finial state, the measurement of
 such channels are very difficult.

Fortunately, having an extra real photon emitted in the leptonic decays, 
the radiative pure leptonic decays 
can escape from the suppression\cite{goldman,eilam,geng}, 
furthermore, as pointed out in Ref.\cite{wang}, with the extra photon 
to identify the decaying pseudoscalar meson $D_s$
in experiment from the backgrounds has advantages, since one more
particle can be detected in the detector.
Although the radiative corrections are suppressed by an
extra electromagnetic coupling constant $\alpha$, it will not be
suppressed by the helicity suppression. Therefore, the
radiative decay may be comparable, even larger than
the corresponding pure leptonic decays\cite{goldman,eilam,geng}.

The radiative pure leptonic decays, theoretically, have infrared 
divergences and will be canceled with those from loop corrections of 
the pure leptonic decays. In all the existing calculation of 
radiative decays\cite{goldman,eilam,geng}, 
this part is ignored, since they do not include
the radiative corrections of the pure leptonic decays. 
In this paper, we are interested in 
considering the radiative decays and the pure leptonic decays with
one-loop radiative corrections together. Since the process
$D_s{\longrightarrow} {{\tau}}{\nu}_{\tau}$ dose not suffer the 
helicity suppression and has a large branching ratio, the corresponding
loop correction(virtual photon) to these process should
 has a considerable larger branching
ratio, at least comparing with the radiative decay, and can not be ignored. 

 The contributions of the radiative decays are 
corresponding to the four diagrams 
in Fig.1. According to the constituent quark model which is formulated 
by Bethe-Salpeter (B.-S.) equation, the amplitude turns out to be
the four terms $M_i (i=1,2,3,4)$:
$$M_1=Tr \left[\int\frac{d^{4}q}{(2\pi)^{4}}\chi(p, q) i 
\left(\frac{G_{F}m_{w}^{2}}{\sqrt{2}
}\right)^{\frac{1}{2}}{\gamma}_{\mu}(1-\gamma_{5})V_{cs}\right]\times $$
$${\frac{ i \left(-g^{\mu\nu}+
\frac{p^{\mu}p^{\nu}}{m_{w}^{2}}\right)}{p^{2}-m_{w}^{2}}}
  i e [(p'+p)_{\lambda}g_{\nu\rho}+
(k-p')_{\nu}g_{\rho\lambda}+(-p-k)_{\rho}g_{\nu\lambda} ] 
{\epsilon}^{\lambda}\times$$

\begin{equation}
\frac{ i \left(-g^{{\rho}{\sigma}}+
\frac{(p-k)^{\rho}(p-k)^{\sigma}}{m_{w}^{2}}\right)}{(p-k)^{2}-
m_{w}^{2}}\overline { \ell }\frac{ig}{2\sqrt{2}}\gamma_{\sigma}
(1-{\gamma}_5)\nu_\ell,
\end{equation}

$$M_2=Tr \left [ \int\frac{d^{4}q}{(2\pi)^{4}}\chi(p, q) i 
\left (\frac{G_{F}m_{w}^{2}}{\sqrt{2}
}\right )^{\frac{1}{2}}{\gamma}_{\mu}(1-\gamma_{5})V_{cs}
\right ]{\frac{ i \left(-g^{\mu\nu}+
\frac{p^{\mu}p^{\nu}}{m_{w}^{2}}\right)}{p^{2}-m_{w}^{2}}}$$

\begin{equation}
 \times\overline { \ell }(-ie)\not\! { \epsilon } \frac{i}{\not\! 
 {k}_{\ell}-m_{\ell}}\frac{ig}{2\sqrt{2}}\gamma_{\nu}(1-{\gamma}_5)\nu_\ell,
\end{equation}
 
$$M_3=Tr \left [ \int\frac{d^{4}q}{(2\pi)^{4}}\chi(p, q) 
i\left (\frac{G_{F}m_{w}^{2}}{\sqrt{2}
}\right )^{\frac{1}{2}}{\gamma}_{\mu}(1-\gamma_{5})V_{cs}
\frac{i}{\frac{m_{s}}{m_{s}+m_{c}}\not\! {p}
+\not\! {q}-\not\! {k}-m_{s}}\left(-i \frac{e}{3}\not\! 
{\epsilon}\right)\right ] $$

\begin{equation}
\times\frac{ i \left(-g^{{\mu}{\sigma}}+
\frac{(p-k)^{\mu}(p-k)^{\sigma}}{m_{w}^{2}}\right)}{(p-k)^{2}-
m_{w}^{2}}\overline { \ell }\frac{ig}{2\sqrt{2}}
\gamma_{\sigma}(1-{\gamma}_5)\nu_\ell,
\end{equation}

$$M_4=Tr \left [ \int\frac{d^{4}q}{(2\pi)^{4}}\chi(p, q)
\left( i \frac{2e}{3}\not\! {\epsilon}\right)
\frac{i}{-(\frac{m_{c}}{m_{s}+m_{c}}\not\! {p}
-\not\! {q}-\not\! {k})-m_{c}} i \left (\frac{G_{F}m_{w}^{2}}{\sqrt{2}
}\right )^{\frac{1}{2}}{\gamma}_{\mu}(1-\gamma_{5})V_{cs}\right] $$

\begin{equation}
\times\frac{ i \left(-g^{{\mu}{\sigma}}+
\frac{(p-k)^{\mu}(p-k)^{\sigma}}{m_{w}^{2}}\right)}{(p-k)^{2}-
m_{w}^{2}}\overline { \ell } \frac{ig}{2\sqrt{2}}\gamma_{\sigma}
(1-{\gamma}_5)\nu_\ell,
\end{equation}
where $\chi(p, q)$ is Bethe-Salpeter wave function of the meson $D_s$; 
$p$ is the momentum of $D_s$; ${\epsilon}, k$ are the polarization 
vector and momentum of the emitted photon.   

 As the $D_s$ meson is a nonrelativistic S-wave
bound state in nature, higher order relativistic corrections
may be small, being the first order approximation for a $S$-wave
state, we ignore $q$ dependence in amplitude and in 
the $D_{s}$ wave-function and write the wave function of the meson $D_s$ as:
$$\int\frac{d^{4}q}{(2\pi)^{4}}\chi(p, q)=
\frac{\gamma_{5}({/}\!\!\! {p}{+m})}{2\sqrt{m}}\psi(0).$$ 
Here $\psi(0)$ is the wave function at origin in the coordinate space,
 and by definitions
it connects to the decay constant $f_{D_s}$:
$$\psi(0)=2\sqrt {m}f_{D_s},$$
where m is the mass of $D_s$ meson. Moreover we note that for 
convenience we take unitary gauge for weak boson to do the calculations
throughout the paper.

There is infrared infinity when performing phase space integral about
the square of matrix element at the soft photon limit. 
It is known that the infrared infinity can be
cancelled completely by that of the 
loop corrections to the corresponding pure leptonic decay $D_s\to \ell\nu$.

If Feynman gauge for photon is taken, 
the amplitude of loop corrections corresponding to the box diagrams (a), (b)\cite{wang} 
can be written as:

$$M_{(2)}(a)={\frac{2}{3}}eA\int\frac{d^{4}l}{(2\pi)^{4}}
\left[\frac{-4i\varepsilon^{\alpha\mu\beta\nu}p_{\alpha}l_{\beta}-4(p_{\mu}l_{\nu}
-{p\cdot l}g_{\mu\nu}+p_{\nu}l_{\mu})+\frac{8m_{c}}{m_{s}+m_{c}}p_{\mu}p_{\nu}}
{l^{2}(l^{2}-2p\cdot l-m_{w}^{2})(l^{2}-{\frac{2m_{c}}{m_{s}+m_{c}}}p\cdot {l}
)[l^{2}-2l\cdot {(p-k_{2})}]}\right]$$
\begin{equation}
\times\overline{\ell}[2(p-k_{2})_{\mu}-\gamma_{\mu}{\not\!   {l}}]
(-\gamma_{\nu})(1-\gamma_{5})\nu_{\ell},
\end{equation}

$$M_{(2)}(b)={-\frac{1}{3}}eA\int\frac{d^{4}l}{(2\pi)^{4}}
\left[\frac{-4i\varepsilon^{\alpha\mu\beta\nu}p_{\alpha}l_{\beta}+4(p_{\mu}l_{\nu}
-{p\cdot l}g_{\mu\nu}+p_{\nu}l_{\mu})-\frac{8m_{s}}{m_{s}+m_{c}}p_{\mu}p_{\nu}}
{l^{2}(l^{2}-2p\cdot l-m_{w}^{2})(l^{2}-{\frac{2m_{s}}{m_{s}+m_{c}}}p\cdot {l}
)[l^{2}-2l\cdot {(p-k_{2})}]}\right]$$
\begin{equation}
\times\overline{\ell}[2(p-k_{2})_{\mu}-\gamma_{\mu}{\not\! {l}}]
(-\gamma_{\nu})(1-\gamma_{5})\nu_{\ell},
\end{equation}
where the $l$, $k_{2}$ denote the momenta of the loop and the
neutrino respectively. These two terms 
also have infrared infinity when integrating out the loop momentum $l$.
 
After doing the on-mass-shell subtraction, the terms 
corresponding to vertex and 
self-energy diagrams (c), (d), (e), (f)\cite{wang} can be written as:
%\begin{center}\begin{eqnarray}
$$M_{(2)}(c+d+e+f)=\frac{ieA}{4\pi^{2}}\overline{\ell}{\not\! p}
(1-\gamma_{5})\nu_{\ell}\times\left[
 ln(4)-\frac{8}{9}+\frac{2}{9}{\frac{m_{s}-m_{c}}{m_{s}+m_{c}}}
ln\left(\frac{m_{s}}{m_{c}}\right)\right.$$%\nonumber\\
 $$+\left(\frac{2}{9}+\frac{8}{9}{\frac{m_{c}}{m_{s}+m_{c}}}\right)
ln\left(\frac{m_{s}+m_{c}}{m_{s}}\right)
 +\left(\frac{8}{9}+\frac{8}{9}{\frac{m_{c}}{m_{s}+m_{c}}}
\right)ln\left(\frac{m_{s}+m_{c}}{m_{c}}\right)$$%\nonumber\\
\begin{equation} +\left.\frac{2}{\varepsilon_{I}}-2\gamma+
ln\left(\frac{4\pi\mu^{2}}{m^{2}}\right)+
 ln\left(\frac{4\pi\mu^{2}}{m_{e}^{2}}\right)\right].
\end{equation}
where $A$ is:
$$A=\frac{f_{D_s}\left(\frac{G_{F}m_{w}^{2}}{\sqrt{2}}\right)^{\frac{1}{2}}
V_{cs}eg}{2\sqrt{2}}=
f_{D_s}\left(\frac{G_{F}m_{w}^{2}}{\sqrt{2}}\right)V_{cs}e.$$

The other loop diagrams(we do not show them) always
have a further suppression factor $m^{2}/{m_{w}}^{2}$
to compare with the loop diagrams we considered, and there is no infrared infinity in
these loop diagrams, we can ignored their contributions safely.
Furthermore we should note that in our 
calculations throughout the paper, the dimensional regularization 
to regularize both infrared and ultraviolet divergences is adopted, 
while the on-mass-shell renormalization for the ultraviolet divergence 
is used.

Detail cancellation of infrared divergence is given in Ref\cite{me}.
Here we simply show the results.
The `whole' leptonic decay branching ratios,
i.e., the sum of the corresponding radiative decay branching ratios  
and the corresponding pure leptonic decay branching ratios
with radiative corrections, and put them in Table (1).
The reason we put the radiative decay 
and the pure leptonic decay with radiative corrections together is to
make the branching ratios not to depend on the experimental 
resolution for a soft photon. For comparison, the 
branching ratios of each pure leptonic decay 
at tree level is also
put in Table (1). The values for the parameters $\alpha=1/132$, 
$|V_{cs}|=0.974$\cite{groom}, $m_{D_s}=1.9686$ GeV, $m_{s}=0.5$ GeV, $m_{c}=1.7$ GeV,
$f_{D_s}=0.24$ GeV\cite{allton} and 
the lifetime $\tau(D_s)=0.469\times10^{-12}s$\cite{groom}.\\

\begin{center}
Table (1) Branching Ratios of the `Whole' and Tree Lever Leptonic Decays \\
\vspace{2mm}
\begin{tabular}{|c|c|c|} \hline
&`whole'&tree\\\hline
 $B_{e}(10^{-5})$ & 2.56&0.0108\\ \hline
 $B_{\mu}(10^{-3})$ & 4.706& 4.605\\ \hline
 $B_{\tau}(10^{-2})$& 4.138&4.489 \\ \hline
\end{tabular}
\end{center}

We can see that, the `whole' decay
branching ratios $Br_{e}$ and $Br_{\mu}$ are 
larger than the corresponding branching ratios of tree lever, 
while the `whole' $Br_{\tau}$ is smaller than the tree lever one. 
It means the contributions
of loop corrections are negative,
the dominate contributions of first order corrections 
to the pure leptonic $D_s$ decays are radiative decays 
 when the lepton is $e$ or $\mu$,  and  
is loop corrections when the lepton is $\tau$. So, the loop contributions
are important for the decays $D_s\to \mu \nu_{\mu}$ 
and $D_s\to \tau \nu_{\tau}$, especially for the later. 
Through Table (1), we obtained that the rediative decay has a branching ratio 
$Br(D_s\to \mu \nu_{\mu}\gamma)> 1.01\times 10^{-4}$ and the loop
correction to $D_s\to \tau \nu_{\tau}$ has a branching ratio 
$Br> 3.51\times 10^{-3}$.

To see the contributions of the radiative decays precisely
we present the radiative decay branching ratios with a cut of the
photon energy, i.e., the branching ratios of the radiative decays 
$D_s\rightarrow l\nu\gamma$ with the photon energy $E_\gamma \geq k_{min}$
as the follows: $k_{min}=0.00001$ GeV, $k_{min}=0.0001$ GeV, 
$k_{min}=0.001 $ GeV, $k_{min}=0.01 $ GeV and $k_{min}=0.1 $ GeV 
respectively in Table (2). We also show the existing results of other methods
 in the same table. 

\begin{center}
Table (2): The Radiative Decay branching ratios 
 with cuts of the photon momentum and the results of Ref\cite{goldman,eilam,geng}\\
\vspace{2mm}
\begin{tabular}{|c|c|c|c|}\hline
$k_{min}$ &$Br_e $&$Br_{\mu}$&$Br_{\tau} $  \\ \hline
GeV  & $10^{-5}$ & $10^{-4} $& $10^{-6} $  \\ \hline
0.00001& 2.552& 4.901& 6.336  \\ \hline
0.0001&  2.552& 3.908& 4.597  \\ \hline
0.001 &  2.551& 2.915& 2.868  \\ \hline
0.01 &   2.549& 1.927& 1.217  \\ \hline
0.1&     2.475& 0.971& 0.727  \\ \hline
Ref\cite{goldman}&   10 &   1  &  $-$   \\ \hline
Ref\cite{eilam}  &   17 & 1.7    &   $-$  \\ \hline
Ref\cite{geng}  &   7.7 &  2.6   &   320  \\ \hline
\end{tabular}
\end{center}

Considering the possible experimental resolutions of photon,
 our prediction of the radiative decay branching ratios $Br(D_s\to e\nu_e\gamma)$ and
$Br(D_s\to \mu\nu_{\mu}\gamma)$ are close to the values in
 Ref\cite{goldman,eilam,geng},
but our prediction of $Br(D_s\to \tau\nu_{\tau}\gamma)$ is much smaller than
the one in Ref\cite{geng}. In our model,
if we using a smaller cut $k_{min}$, then obtained a larger
$Br(D_s\to \ell\nu_{\ell}\gamma)$, 
 because the decay widths depend
on $Log(k_{min})\cite{me}$, the change of branching ratios
 will be not so much on the selection of $k_{min}$, we can
see this in Table (2), and for another example, if $k_{min}=1.0\times 10^{-10}$ GeV, 
then we obtain $Br(D_s\to e\nu_{e}\gamma)=3.59\times 10^{-5}$, 
$Br(D_s\to \mu\nu_{\mu}\gamma)=1.38\times 10^{-3}$, 
$Br(D_s\to \tau\nu_{\tau}\gamma)=2.11\times 10^{-5}$, but so small a $k_{min}$,
it is very difficult in experiment.
 We can conclude that the best radiative decay channel is easy
 to search in experiment is $D_s\to \mu\nu_{\mu}\gamma$. 

For the convenience to compare with experiments, we present the photon
spectrum of the radiative decays in Fig.2 and Fig.3.   
In addition, we should note that the widths are quite sensitive to 
the decay constant $f_{D_s}$, and are sensitive to
 the values of the quark masses 
$m_{s}$ and $m_c$.

%\begin{}

\begin{figure}\begin{center}
   \epsfig{file=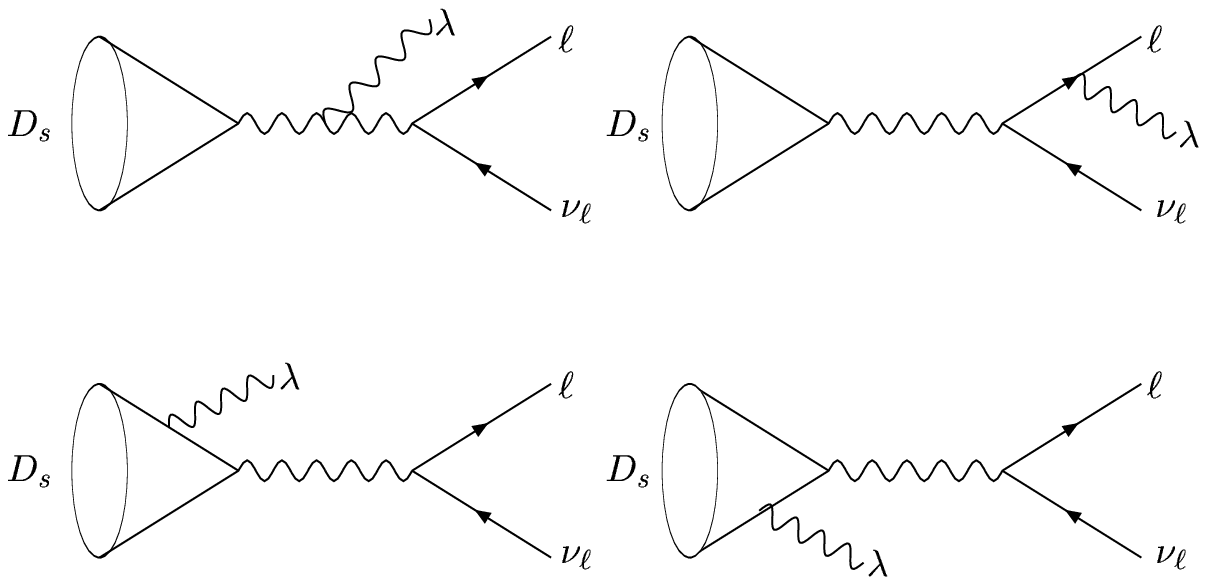, bbllx=157pt,bblly=497pt,bburx=505pt,bbury=683pt,
width=8cm,angle=0}
\caption{\bf Diagrams for $D_s\longrightarrow \ell \nu \gamma $.}
%\label{fig}
\end{center}
\end{figure}

\begin{figure}\begin{center}
   \epsfig{file=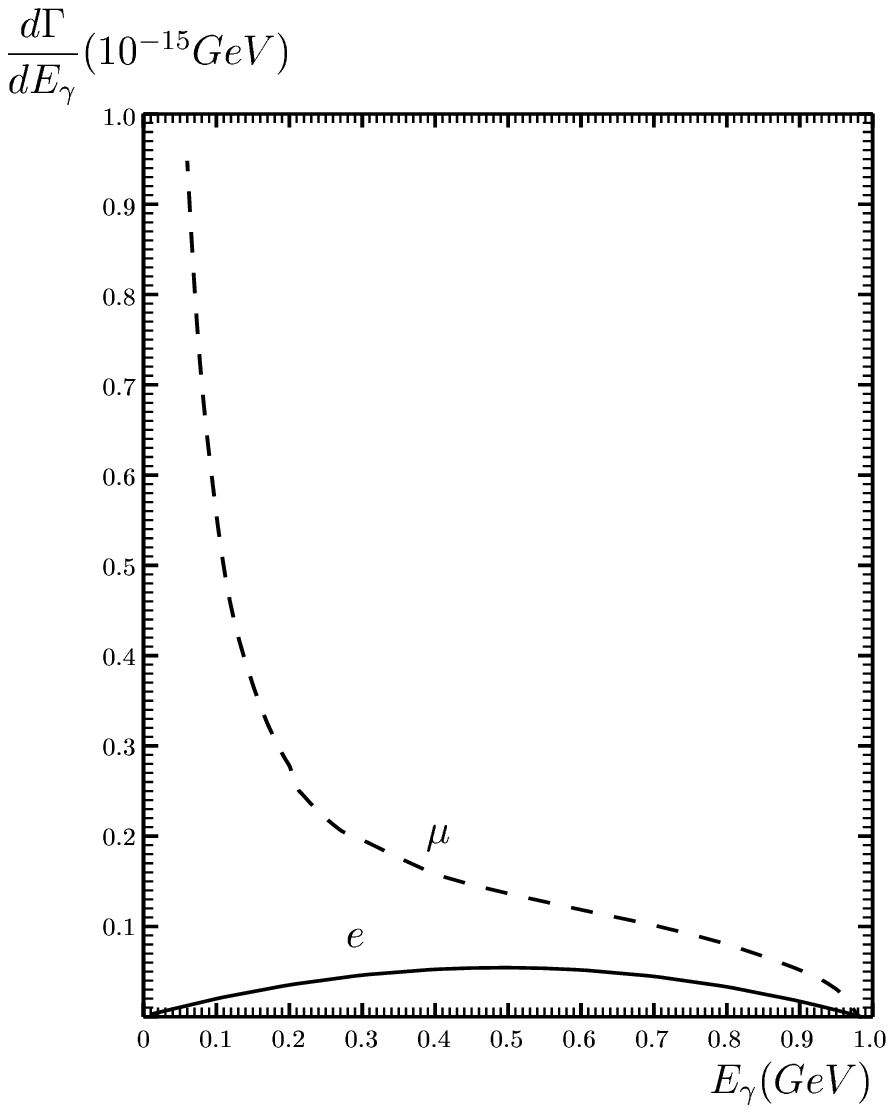, bbllx=216pt,bblly=365pt,bburx=450pt,bbury=680pt,
width=7cm,angle=0}
\caption{\bf Photon energy spectra of radiative decays  $D_s\longrightarrow\ell{\nu_{\ell}}
\gamma(\ell=e, \mu)$.}
%\label{fig}
\end{center}
\end{figure}
\begin{figure}\begin{center}
   \epsfig{file=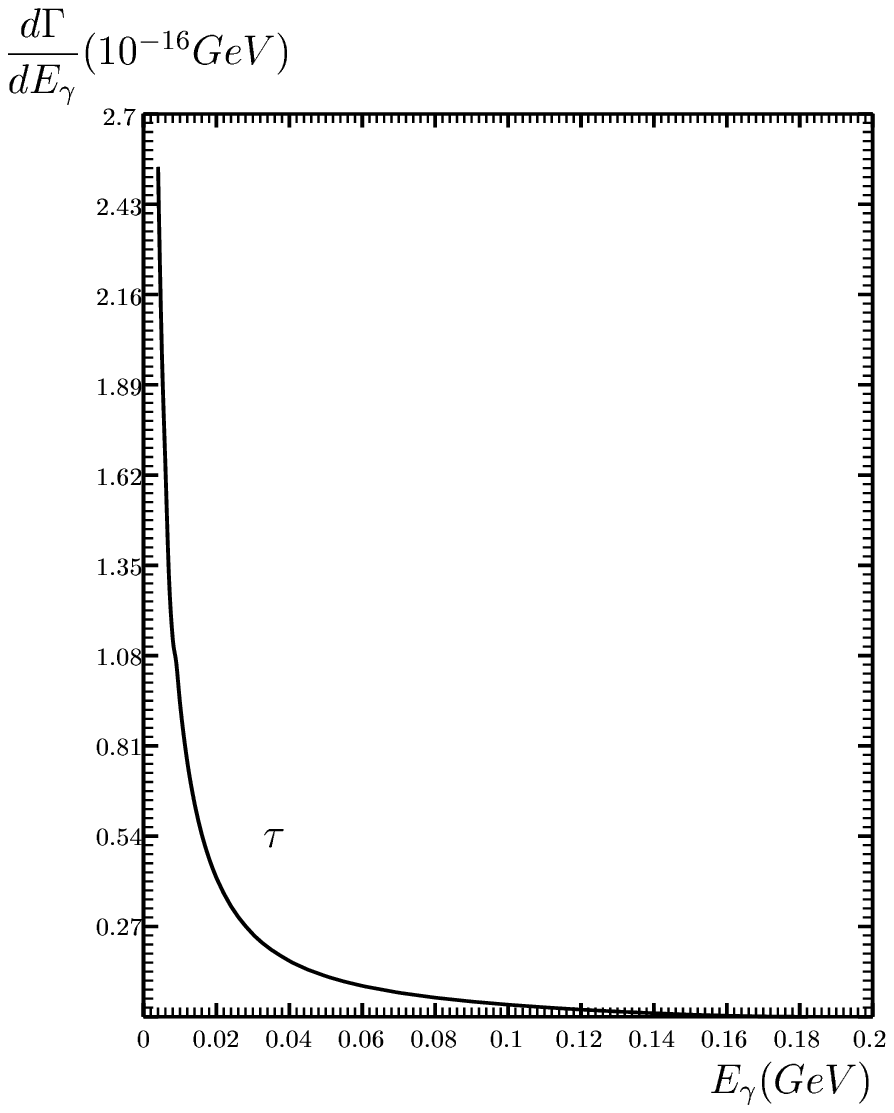, bbllx=216pt,bblly=365pt,bburx=453pt,bbury=664pt,
width=7cm,angle=0}
\caption{\bf Photon energy spectra of radiative decays 
 $D_s\longrightarrow\tau{\nu_{\tau}}
\gamma$.}
%\label{fig}
\end{center}
\end{figure}

\end{document}